# Nanowire Array Breath Acetone Sensor for Diabetes Monitoring

*Shiyu Wei, Zhe Li, Krishnan Murugappan, Ziyuan Li, Mykhaylo Lysevych, Kaushal Vora, Hark Hoe Tan, Chennupati Jagadish, Buddini I Karawdeniya\*, Christopher J Nolan, Antonio Tricoli\*, and Lan Fu\**

S. Wei, Z. Li, Z. Y. Li, H. H. Tan, C. Jagadish, B. I. Karawdeniya, L. Fu
Australian Research Council Centre of Excellence for Transformative Meta-Optical Systems, Department of Electronic Materials Engineering, Research School of Physics, The Australian National University, Canberra, ACT 2600, Australia

M. Lysevych, K. Vora
Australian National Fabrication Facility, The Australian National University, Canberra, ACT 2600, Australia

K. Murugappan
Commonwealth Scientific and Industrial Research Organisation (CSIRO), Mineral Resources, Private Bag 10, Clayton South, Victoria, 3169, Australia
Nanotechnology Research Laboratory, Research School of Chemistry, College of Science, The Australian National University, Canberra, ACT 2600, Australia

A. Tricoli
Nanotechnology Research Laboratory, Faculty of Engineering, The University of Sydney, Camperdown 2006, Australia
Nanotechnology Research Laboratory, Research School of Chemistry, College of Science, The Australian National University, Canberra, ACT 2600, Australia

C. J. Nolan
School of Medicine and Psychology, College of Health and Medicine, The Australian National University, Canberra, ACT 2600, Australia
Department of Diabetes and Endocrinology, The Canberra Hospital, Garran, ACT 2605, Australia

E-mail: buddini.karawdeniya@anu.edu.au; antonio.tricoli@anu.edu.au; lan.fu@anu.edu.au

Keywords: Diabetic ketoacidosis, acetone sensor, InP nanowires, chitosan, breath test prototype

Abstract

Diabetic ketoacidosis (DKA) is a life-threatening acute complication of diabetes in which ketone bodies accumulate in the blood. Breath acetone (a ketone) directly correlates with blood ketones, such that breath acetone monitoring could be used to improve safety in diabetes care. In this work, we report the design and fabrication of a chitosan/Pt/InP nanowire array based chemiresistive acetone sensor. By implementing chitosan as a surface functionalization layer and a Pt Schottky contact for efficient charge transfer processes and photovoltaic effect, self-powered, highly selective acetone sensing has been achieved. This sensor has an ultra-wide detection range from sub-ppb to >100,000 ppm levels at room temperature, incorporating the range from healthy individuals (300-800 ppb) to those at high-risk of DKA (> 75 ppm). The nanowire sensor has been further integrated into a handheld breath testing prototype, the Ketowhistle, which can successfully detect different ranges of acetone concentrations in simulated breath. The Ketowhistle demonstrates immediate potential for non-invasive ketone testing and monitoring for persons living with diabetes, in particular for DKA prevention.

1. Introduction

With the number of individuals diagnosed with diabetes mellitus predicted to increase to 643 million by 2030 worldwide, the demand for rapid, sensitive, and low-cost health monitoring and diagnostic technologies is growing exponentially.[1] All of those living with type 1 diabetes and an increasing number with type 2 diabetes, particularly those managed with sodium glucose co-transporter inhibitor 2 (SGLT2) glucose lowering medications, are at risk of diabetic ketoacidosis (DKA), a serious and life-threatening condition.[2] DKA occurs as a consequence of insulin deficiency, resulting in uncontrolled release of fatty acids from the fat and excessive production of ketones from these fatty acids in the liver.[3] Ketones in blood are either metabolised by the body for energy, or excreted from the body through urine, sweat, and exhaled breath. In urine and sweat, ketones are normally present as 3-β-hydroxybutyrate (3β-HB) and acetoacetate (AcAc) respectively, while acetone is a small volatile ketone in the breath.[4] Due to the seriousness of DKA, monitoring is advised, particularly in situations of increased risk (e.g. in sick day management). Current clinical detection methods largely rely on blood and urine assays, but breath ketone monitoring is not established.[5] Finger-prick tests using ketone strips are the most common method to monitor ketone levels in the body; however the invasive nature has led to low levels of compliance, especially in children and younger individuals. Adding to poor compliance with blood ketone testing is the limited shelf life of the blood ketone testing strips and their cost. Urine testing is usually not recommended, as it is inconvenient causing issues of compliance, and less accurate.

Considering the seriousness of DKA, the increasing number of people at risk with the advent of the SGLT2 inhibitors, and current issues with using blood and urine for ketone monitoring which are particularly challenging for children, innovations in how ketone levels are monitored are needed. A rapid and non-invasive breath sensing methodology could deliver a solution. Acetone concentrations in breath have a direct correlation with blood ketone levels,[6] making acetone in breath an ideal biomarker for use in DKA detection in point of care and self-use diagnostics. The exhaled acetone level of individuals with type 1 and 2- diabetes when well tends to be a little higher than in non-diabetic individuals (<1 ppm), but are usually <2 ppm.[7] In DKA, however, breath acetone ranges from >75 ppm to 1200 ppm.[8] Overlap with the lower end of the DKA range can occur with prolonged fasting and the use of keto-diets for weight loss or management of epilepsy. To ensure practical relevance, the designed sensors should have a rapid response rate, high selectivity, and a sufficient lower detection limit (sub-ppm) and upper detection limit (about 2000 ppm). These requirements are highly challenging, especially considering the high humidity, temperature fluctuations, interference of contaminants and alternance of gases in human breath.[6, 9] Analysing exhaled breath samples with point-of-care or portable gas sensors is a new frontier in healthcare and medicine,[10] which is promising and undergoing rapid development to build better diagnostic tools. However, so far, only a few breath tests have been extended to clinical applications,[11] and these applications rarely use portable, self-use, real-time diagnostic devices, highlighting the demand for relevant research.

Chemiresistive gas sensors are promising for breath analysis applications due to their superior sensitivity, material designability, low power consumption, compact size, and easy integration into handheld devices.[12] Recently, chemiresistive acetone sensors with ppm level sensitivity have been demonstrated using metal-oxides[13] and 2D materials.[14] For both types of sensors, acetone sensing relies on forming chemisorbed oxygen species from ambient oxygen molecules that trap electrons from the sensing materials. To form chemisorbed oxygen species, e.g., $O_2^-$, $O^-$, and $O^{2-}$, most of these devices must operate at high working temperatures (>300 °C) and/or be activated by light illumination, resulting in high power consumption and safety related issues.

Similar sensing reactions with other volatile organic compounds (VOCs) have also been observed, resulting in limited selectivity of many reported acetone sensors.[15]

Chitosan-based acetone sensing at low working temperatures (25 - 30 °C) has also been reported.[16] Chitosan is a low-cost, renewable, and eco-friendly biopolymer[17] with superb water solubility, biodegradability, renewability, antimicrobial activity, biocompatibility, and adsorption properties.[16b] The amino group (-NH$_2$) in chitosan can interact with the carbonyl groups of acetone, contributing to a high acetone specificity.[18] However, as the active layer of acetone sensors, chitosan shows inadequate selectivity and/or sensitivity to fulfil the clinical requirements.[16] On the other hand, recently it has been shown that InP nanowire (NW) arrays are a promising gas-sensing material platform. By carefully tailoring the NW array geometry, device structure, and fabrication design, NO$_2$ gas sensors[19] with sub-ppb level sensitivity, rapid response time, room temperature and self-powered operation[27] can be achieved. However, it has also been found that InP NWs only exhibit high sensitivity towards NO$_2$ due to its strong oxidizing nature, but do not have high sensitivity to weak oxidizing or reducing gases such as CO$_2$ and acetone.

Here, we report for the first time the design and fabrication of a new type of high-performance self-powered InP NW array acetone sensor, using chitosan as a surface functionalization layer to achieve selective detection of acetone molecules. By implementing a Pt Schottky contact layer, highly sensitive and self-powered operations have been realized through efficient charge transfer processes and photovoltaic (PV) effect owing to the ideal Pt/InP band alignment. The acetone-sensing mechanism was studied through numerical simulations and experimental investigations, revealing an oxygen-facilitated two-step charge transfer process enabling acetone redox reaction. Finally, the NW sensor has been integrated into a portable electronic setup, the Ketowhistle, as a prototype breath testing device that successfully detects simulated exhaled breath containing different ranges of acetone concentrations, demonstrating great promise as a low-cost and user-friendly breath analysis device for ketone monitoring in people living with diabetes and in particular, DKA prevention.

2. InP NW Acetone Sensor Fabrication

The 400 × 400 μm InP NW arrays were grown by a bottom-up method based on the selective area metal-organic chemical vapor deposition (SA-MOCVD) technique,[26] with a diameter of 50 - 60 nm, pitch size of 600 nm and length of ~ 4 μm (details in Supporting Information (SI),

Figure S1). The growth condition was optimized to produce thin NWs with pure wurtzite crystal structure due to dominant axial growth versus radial growth.[20]

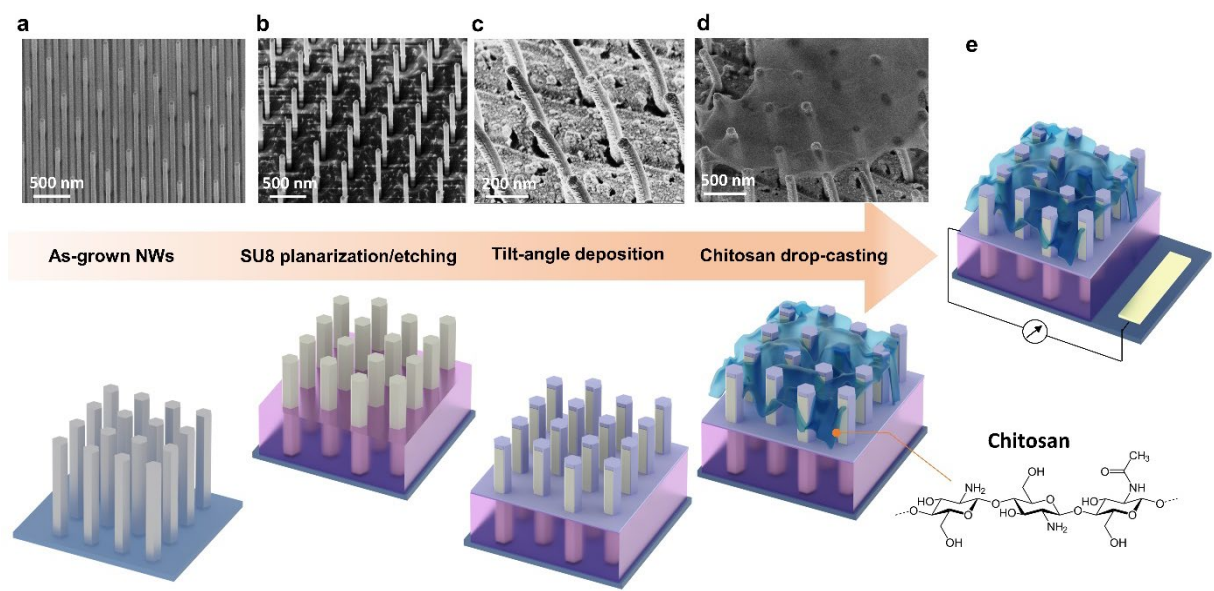

**Figure 1. The fabrication process of the chitosan/Pt/InP NW array acetone sensor.** Scanning electron microscope (SEM) images in the top panel with the corresponding schematics in the bottom panel represent the sensor device fabrication processes, including **a**, the as-grown NWs; **b**, SU8 planarization and etching to expose the top portion of the NWs; **c**, tilt-angle deposition of Pt electrode, causing a slight NW bending due to the small NW diameter; **d**, drop-casting of chitosan for surface functionalization in 45° tilt-angle imaging; and **e**, the Ti/Au bottom contact fabrication next to the NW array for electrical connection.

The as-grown InP NW arrays were then fabricated into the acetone sensor following the process illustrated in **Figure 1** (see Method for details). Firstly, the SU8-5 photoresist was spin-coated to planarize the NW array, followed by oxygen plasma etching to uniformly expose the top segments of the NWs (Figure 1b). This provides mechanical stability of the thin NWs and electrical isolation between the exposed NW tops and the substrate. Then, the tilt-angle deposition method[28] (Figure 1c) was applied to deposit a thin layer (< 50 nm) of Pt contacting one side of the NW. This enables all the NWs in the array to be electrically connected, meanwhile, allowing the other side of NW surface un-deposited for light absorption and gas molecule interaction. Then, a diluted chitosan solution was drop-casted over the NW/Pt array to form a parafilm-like layer, functionalizing the exposed NW surface, as shown in Figure . 1d. Finally, a Ti/Au contact was fabricated on top of the InP substrate next to the NW array, to complete the fabrication of the acetone sensor based on bottom up grown NW array, i.e, the b-chitosan/Pt/InP NW sensor (Figure 1e).

3. Result and Discussion

3.1. Acetone sensor performance characterization

After the sensor fabrication, the sensing measurements are performed using a custom-made gas sensing system comprised of a Linkam chamber with a sample stage and Au probes (for contacting sample electrodes), multiple gas inlets connected with mass flow controllers, a solar simulator and gas cylinders (**Figure 2**a). As shown in Figure 2b, the electrical properties of the sensor were characterized by standard dark/light current-voltage (I-V) measurements, exhibiting a typical Schottky diode characteristic. When illuminated under a solar simulator @AM1.5 at a power of 100 mW·cm$^{-2}$ (1 sun), the device displays a short circuit current ($I_{SC}$) of $1.4 \times 10^{-7}$ A (increase from $< 10^{-9}$ A in dark) and an open-circuit voltage ($V_{OC}$) of 80 mV, indicating the capability of For the sensing measurement, the baseline $I_{SC}$ signal was measured under 1 sun illumination with constant airflow and zero external bias (self-powered). Before gas injection, the sensor was placed in the chamber under light illumination and constant simulated airflow to generate steady-state baseline $I_{SC}$. As acetone gas is injected, $I_{SC}$ increases and then reaches saturation. The variation of saturated and baseline $I_{SC}$ is calculated as the sensing response (R), as defined by equation (1) in Methods. When acetone gas was switched off and airflow resumed, $I_{SC}$ returned to the initial base level.

The sensing performance of the b-chitosan/Pt/InP NW sensor at low acetone concentrations was investigated from 0.4 ppb to 10 ppm. As shown in Figure 2c-f, the NW sensor exhibits a strong concentration-dependent sensing response, demonstrating high sensitivity at both ppb and ppm levels. The linear fitting results in Figure 2g illustrate R as a function of acetone concentration C (ppb) with a slope that defines the device sensitivity S = R/C(ppb).[21] The slope in the higher concentration range (100 - 1000 ppb) is 0.032 %/ppb, smaller than that in the 0.4 - 10 ppb range (0.48 %/ppb). The decreased sensitivity at a higher concentration range results from the charge-release saturated for increasing analyte concentration, given the finite number of active surface sites[22]. Based on the linear fitting result, the limit of detection (LOD) of our acetone sensor is calculated to be 0.18 ppb (detail in Method), indicating an excellent sensitivity, which satisfies the clinical requirement (~ 1 ppm) for lower level breath acetone detection.[13] It is worth noting that our sensor also features a rapid response time ($T_{res}$: ~25 s) and recovery time ($T_{rec}$: ~39 s), as shown in Figure S2 (SI), promising for real-time diagnosis. The reproducibility of the b-chitosan/Pt/InP NW sensor was tested by repeating 16 cycles of acetone sensing measurement (SI, Figure S3). A relatively consistent R value is obtained with the maximum standard deviation of only 5.45%, indicating excellent stability.self-powered operation.

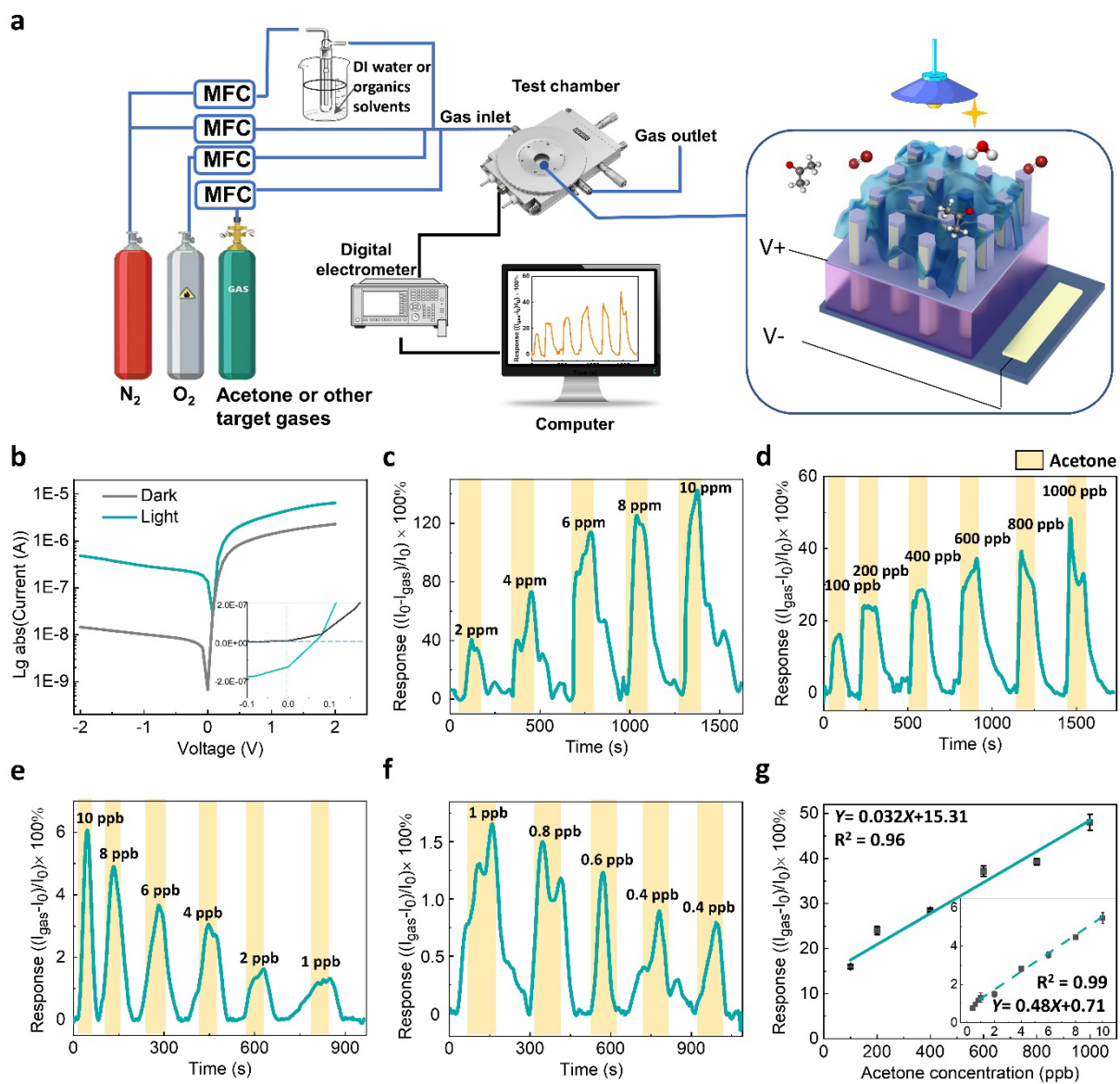

**Figure 2. Acetone sensing performance of the b-chitosan/Pt/InP NW sensor. a**, Schematic of the lab gas sensing setup for acetone sensing measurements. **b**, I-V curves measured from the sensor device under dark/light conditions. Inset: zoom-in plot of the I-V curve for displaying $I_{SC}$ and $V_{OC}$ **c**, Time-dependent sensing response measured from short-circuit current under light illumination for different ranges of acetone concentration: **c,** 2 – 10 ppm, **d**, 100 - 1000 ppb, **e**, 1 - 10 ppb, and **f** 0.4 - 1 ppb. The yellow strips indicate acetone exposure. **g**, Concentration vs response curve for the acetone concentration range of 100 - 1000 ppb and 0.4 - 10 ppb (inset) respectively.

Apart from the high sensitivity, the selectivity of the b-chitosan/Pt/InP NW sensor is also critical because there are numerous VOCs in human breath with concentrations varying from ppt to ppm levels. A selective sensor for breath ketone testing is required to distinguish acetone from these interfering VOCs. In this case, the sensitivity of the NW sensor to a series of VOCs and atmospheric gases was measured at a concentration of 1 ppm, including acetone ($CH_3COCH_3$), methyl nitrides ($CH_3NO_2$), ethanol ($C_2H_5OH$), propane ($C_6H_6$), and 1% carbon dioxide ($CO_2$). As shown in **Figure 3**a, the sensing responses from these gases were less than 5% of that from acetone (49 ± 2%), indicating an excellent sensitivity towards acetone. In particular, ethanol is the most common interference in breath; however the acetone selectivity over ethanol interference ($R_{acetone} / R_{ethanol}$) of our sensor is > 40, much higher than that obtained in previous studies.[23] The bubbler setup (Figure 2a) was further used to evaporate acetone and 2-butanone solvents at concentrations 10,000 times higher than that used in sensing measurement in Figure 2. As shown in Figure 3b, the sensor maintained a positive sensing response correlation with the concentration from 0.57% (5,700 ppm) to 11.1% (160,500 ppm), indicating a wide dynamic range within which the clinically important range from sub-ppm (healthy person) to 1000 ppm (DKA patient) will easily fit. It is worth noting that even though both acetone and 2-butanone molecules are small ketones, our NW sensor only demonstrated a high sensitivity to acetone. A possible reason is that the activation energy required for the 2-butanone sensing reaction is higher than that required for acetone.[24] This selectivity is of great significance for clinical breath analysis since 2-butanone is a small ketone molecule (like acetone) and a biomarker present in breath (i.e., lung cancer), and the capability to distinguish it from acetone will greatly reduce the risk of false positives.[10a]

Another major challenge in breath analysis is the water vapor in the breath, which could impact the sensor behaviour by saturating the sensor and decreasing the material sensitivity to the target gas. The effect of humidity on the acetone sensing performance was characterized on b-chitosan/Pt/InP NW devices by the same bubbler setup to create a humid condition. As shown in Figure 3c, compared to the dry air condition, the sensing response decreased by 10 ± 2 %, 30 ± 6 %, and 50 ± 12 % under a relative humidity (RH) of 20%, 50%, and 65%, respectively (detailed in Method and Table S1). Nevertheless, from the sensing response slope plotted in Figure 3d, the device still maintains a consistent sensitivity to acetone, indicating that the sensitivity of this device was not compromised in a humid environment. The cross-sensitivity to interrupting molecules that may be present in exhaled breath, such as ethanol and $CO_2$, was also measured under different RH conditions separately, as shown in Figure 3e, confirming the high selectivity towards acetone, with a response more than ten times higher than those from the interfering gases under high RH conditions.

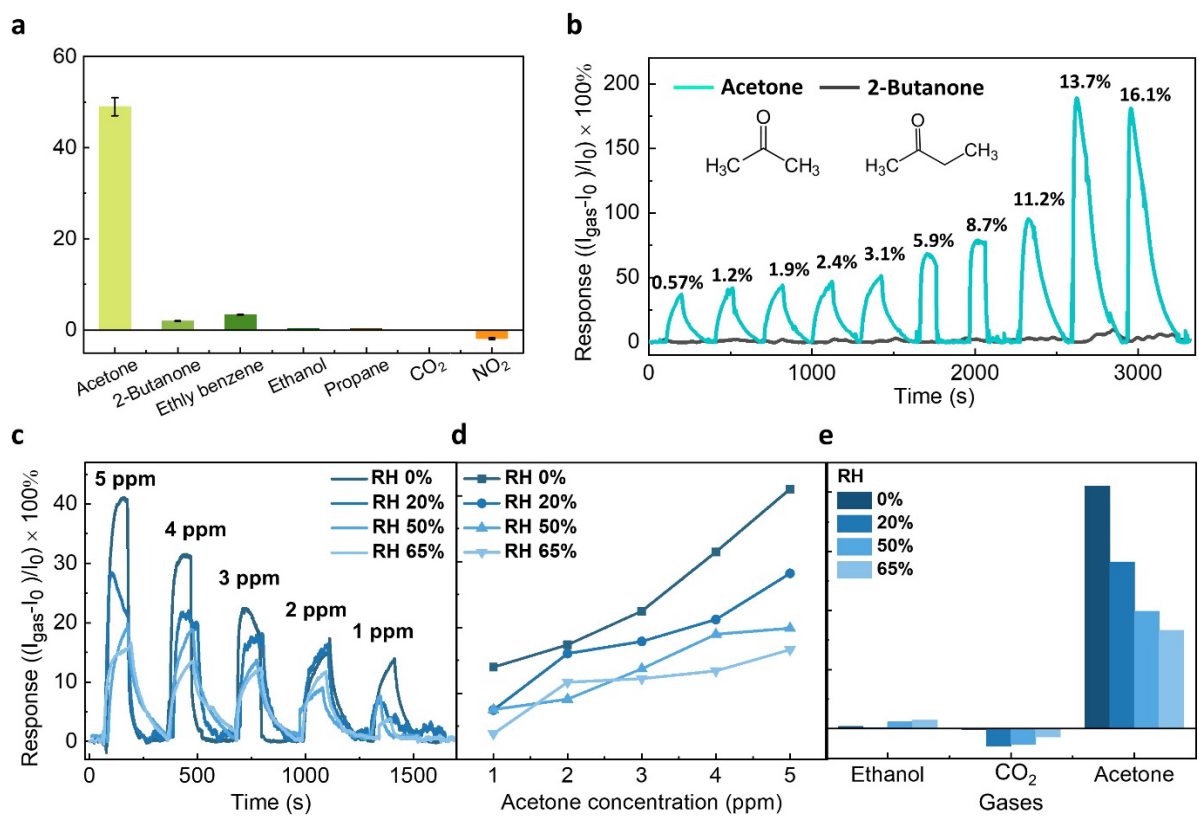

**Figure 3. Selectivity and humidity test of the b-chitosan/Pt/InP NW sensor**. **a**, Selectivity measurements performed for different gases, including acetone, 2-butanone, ethyl benzene, ethanol, propane, and $NO_2$ under a concentration of 1 ppm, and $CO_2$ of 1%, with the standard deviation shown as error bars derived following 10 cycles of sensing measurements. **b**, Time-dependent sensing response of acetone and 2-butanone with concentrations ranging from 0.57% (5,700 ppm) to 16.05% (160,500 ppm). **c**, **d**, Sensing response to the acetone concentration of 1-5 ppm under the RH levels of 0%, 30%, 50%, 65%. **e**, Gas sensing selectivity measurement to 1 ppm acetone, ethanol and 1 % $CO_2$ under different RH conditions.

The key performance parameters of the b-chitosan/Pt/InP NW sensor and those of other chemiresistive acetone sensors reported in recent years are summarized in Table S2 (SI). The NW sensor has shown superior performance in all key metrics, including sub-ppb level detection, rapid response/recovery speed, and an ultra-broad dynamic range covering different types of diabetes and metabolic conditions, satisfying for the basic clinical requirements for breath testing in different healthcare settings. Also, for the first time, the self-powered, room temperature operation has been achieved by our NW sensor, which will effectively address the power consumption and safety-related issues.

Furthermore, as the bottom-up technology based b-chitosan/Pt/InP NW sensor relies on specialized epitaxial growth equipment and technique, which is of high cost, we also developed an alternative top-down etching approach to fabricate InP NW arrays and the corresponding t-chitosan/Pt/InP NW sensors[25] (Figure S4, SI,). This approach is low-cost and compatible with the well-established complementary metal-oxide semiconductors (CMOS) processes.[26] The details on device processing and sensing performance testing are provided in Methods and Figure S5. Compared to the bottom-up device, the t-chitosan/Pt/InP NW sensor has a slightly different NW morphology (i.e., size, shape, and surface roughness). It exhibits shorter $T_{res}$ (~5 s) and $T_{rec}$ (~7 s), higher humidity tolerance, light intensity fidelity, selectivity, and ease of fabrication, despite slightly lower sensitivity, as shown in Figure S6.

3.2. Investigation of the chitosan/Pt/InP NW acetone sensing mechanism

To understand the sensing mechanism of the chitosan/Pt/InP NW acetone sensor, a series of controlled experiments (**Figure 4**a, b) and simulation studies (Figure 4c, f) have been performed. Contrary to previous studies in which chitosan acts as the active sensing layer,[16] in our study the chitosan layer was drop-casted on top of the NW array as the surface functionalization layer and not electrically connected to the sensing circuit. To clarify the effect of chitosan on acetone sensing, a similar device without the chitosan layer was fabricated and used as a reference sample. Figure 4a compares the baseline $I_{SC}$ of the Pt/InP NW device with/without chitosan under light illumination. Compared with the device without chitosan, the chitosan-coated device has a much lower baseline light current. We ascribed it to the ability of chitosan to absorb and release oxygen.[27] The presence of chitosan effectively absorbs $O_2$ from the carrier gas flow of simulated air (details see Method), trapping $O_2$ molecules at the chitosan/InP interface, as denoted in equation (1). This would facilitate $O_2$ ionization (equation (2)) by the photogenerated electrons that drift to the InP NW surface due to the built-in electric field formed with the Pt/InP Schottky contact (see detailed discussion later). This leads to a decreased baseline current ($I_{SC}$) of the chitosan-coated device, as shown in Figure 4a.

$O_{2(gas)} \leftrightarrow O_{2(ads)}$ (1)

$O_{2(ads)} + e^- \leftrightarrow O_2^-_{(ads)}$ (2)

$CH_3COCH_{3\ (ads)} + 4O_2^-_{(ads)} \rightarrow 3CO_2 + 3H_2O + 4e^-$ (3)

In contrast, the device without chitosan had much less $O_2$ adsorption on the InP NW surface, and thus, a negligible baseline $I_{SC}$ change was observed. As the main mechanism for most chemiresistive acetone sensors, these oxygen ions are the indispensable reactant to enable a redox reaction with acetone, producing $CO_2$, $H_2O$ and electrons,[15a, 28], as depicted in equation (3). Therefore, it is clear that chitosan plays a critical role in surface functionalization, not only through its -$NH_2$ groups to enhance acetone adsorption[16b] and selectivity, but also in improving $O_2$ absorption. As can be seen from Figure 4b, with the simulated air, the acetone sensing response obtained from the chitosan-coated device (>120% at 10 ppm) is significantly higher than the device without chitosan (< 5% at 10 ppm). However, under the $O_2$ eliminated condition (only pure $N_2$ as the carrier gas for sensing measurement, details see Method), the sensing response of the chitosan-coated device decreased from > 120% (with $O_2$) to < 15% (without $O_2$), as shown in the middle panel of Figure 4b, confirming the important involvement of the $O_{2(ads)}$ and ionic oxygen species like $O_2^-$ in the acetone sensing reaction.

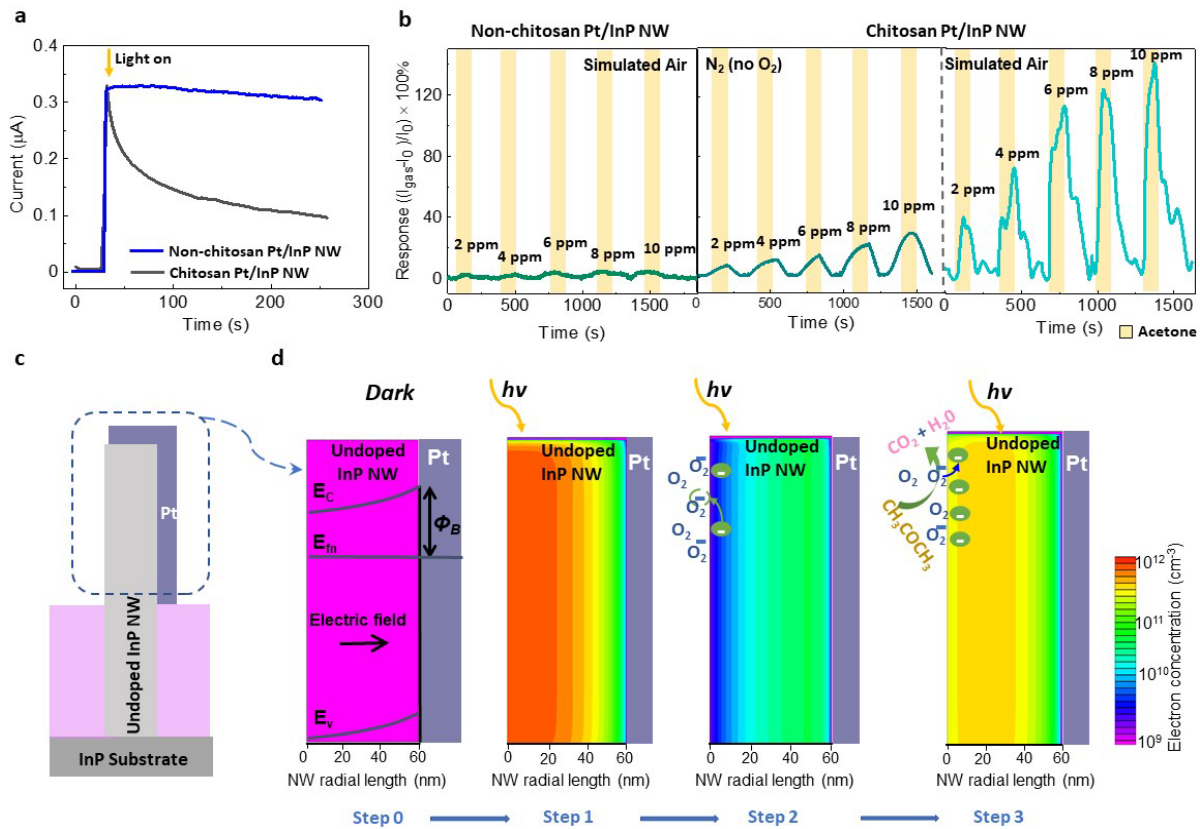

**Figure 4. Acetone sensing mechanism investigation. a**, Temporal short-circuit current ($I_{SC}$) response of the Pt/InP NW device with and without chitosan under light illumination. **b**, Time-dependent acetone sensing response of the sensor with and without chitosan (under low oxygen conditions and in simulated air). **c**, Schematic of an InP NW with Pt contact used for the simulation study. **d**, Band energy diagram of the structure with the simulated electron concentration, corresponding to the various critical condition/steps of acetone sensing process. $E_C$, $E_v$, and $E_{fn}$ represent the conduction band, valence band, and Femi level, respectively.

A simulation study by COMSOL Multiphysics was performed to investigate the effect of Pt/NW Schottky contact by calculating the energy band diagram of the device and electron concentration distribution within a single NW (Figure 4c, d). As shown in Figure 4d, under the dark condition (Step 0), the Pt layer forms a Schottky contact on the InP NW side wall. Due to the alignment of the Fermi level between the InP NW and Pt contact, an electron depletion layer is first created that drastically decreases the electron concentration within the NW due to the small NW diameter (50 - 60 nm). Upon light illumination, the built-in electric field generated by the Schottky contact drives photo-generated electrons towards the NW surface (Step 1), resulting in surface accumulation of electrons. This facilitates the ionization of adsorbed $O_2$ on the NW interface (Step 2), as enhanced by chitosan functionalization and expressed in equation (1) and (2). The enhancement of $O_2$ ionization enables the acetone reduction reaction (equation (3)) to release electrons back to the InP NW and significantly increases the carrier concentration in the NW (Step 3).

Based on above, it is evident that the sensing response is closely related to the two-step electron transfer processes and the resultant modulation of carrier concentration in the InP NWs. The Schottky contact, NW diameter, and doping concentration must be properly designed to achieve the optimum performance, as shown and discussed in Figure S7 and S8, SI.

3.3. Portable breath sensor (Ketowhistle) development, calibration, and testing

To demonstrate the applicability of the chitosan/Pt/InP NW array acetone sensor for point-of-care application, a prototype portable device has been demonstrated for exhaled breath testing[29] by integrating the nanowire sensor onto a handheld apparatus, named Ketowhistle, as shown in **Figure 5**a. Considering the fabrication cost and scalability issue, the top-down fabricated device was chosen for integration into the Ketowhistle.

The handheld Ketowhistle apparatus contains a t-chitosan/Pt/InP NW sensor, a commercial $CO_2$ sensor (to ensure end-tidal expired breath most reflective of blood acetone concentrations[30] is tested), a signal-processing circuitry, and an OLED screen for result display (electrical diagram provided in SI, Figure S9 and Table S3). The NW acetone sensor is illuminated by a low-power red LED, which is placed right on top of the sensor to ensure stable illumination and sensing response, as shown in Figure 5a. Considering the wide acetone concentration range corresponding to a variety of physiological states, as summarized in Figure 5b,[31] wide concentration range calibration of the Ketowhistle for the breath-based acetone analysis was performed using the laboratory based gas sensing system. Figure 5c presents the Ketowhistle calibration for acetone concentration ranging from 0.1-1000 ppm to cover the entire breath acetone spectrum. The measured electrical signal readout from the Ketowhistle exhibits two distinct regions, with a slope of 6.11 mV/ppm and 0.013 mV/ppm corresponding to two ranges of acetone concentration, i.e., Range 1 (0.1 - 10 ppm) and Range 2 (50 - 500 ppm), respectively (calibration data see Figure S10, SI). Within each range, there is a linear correlation between the electrical readout and the acetone concentration, which is consistent with the lab-based non-integrated device measurement in Figure 2. It is worth noting that, as indicated in Figure 5b, the low acetone concentration Range 1 covers the range that is reflective of normal eating patterns, whereas Range 2 indicates significant ketosis with high risk of DKA. For persons with established diabetes, a device giving a digital readout of concentrations would be ideal in the management of sick days, as the trend in levels will indicate improvement or deterioration. These two ranges could be applied to a "traffic light" system very useful in devices used for diabetes screening (e.g. in primary care clinics) with 0.1-10 ppm shown as green (indicative of low risk of DKA), 50 to 500+ ppm shown as red (indicative of high risk of DKA), and 10-49 ppm shown as yellow (a need for caution, there may a risk of progressing to DKA). Also, the approximate linear correlation of the output voltage signal with a wide range of acetone concentration allows this device to be applied beyond diabetes management, such as the monitoring of ketogenic diets in weight loss plans, and monitoring of nutrient status in atheletes.[32] Although ketones are not monitored routinely in women with GDM, there is concern for the risk of ketosis affecting pregnancy outcomes if carbohydrates are too severely restricted. Thus, a simple to use device for ketone monitoring, such as the Ketowhistle, may also have a future role in pregnancy care.

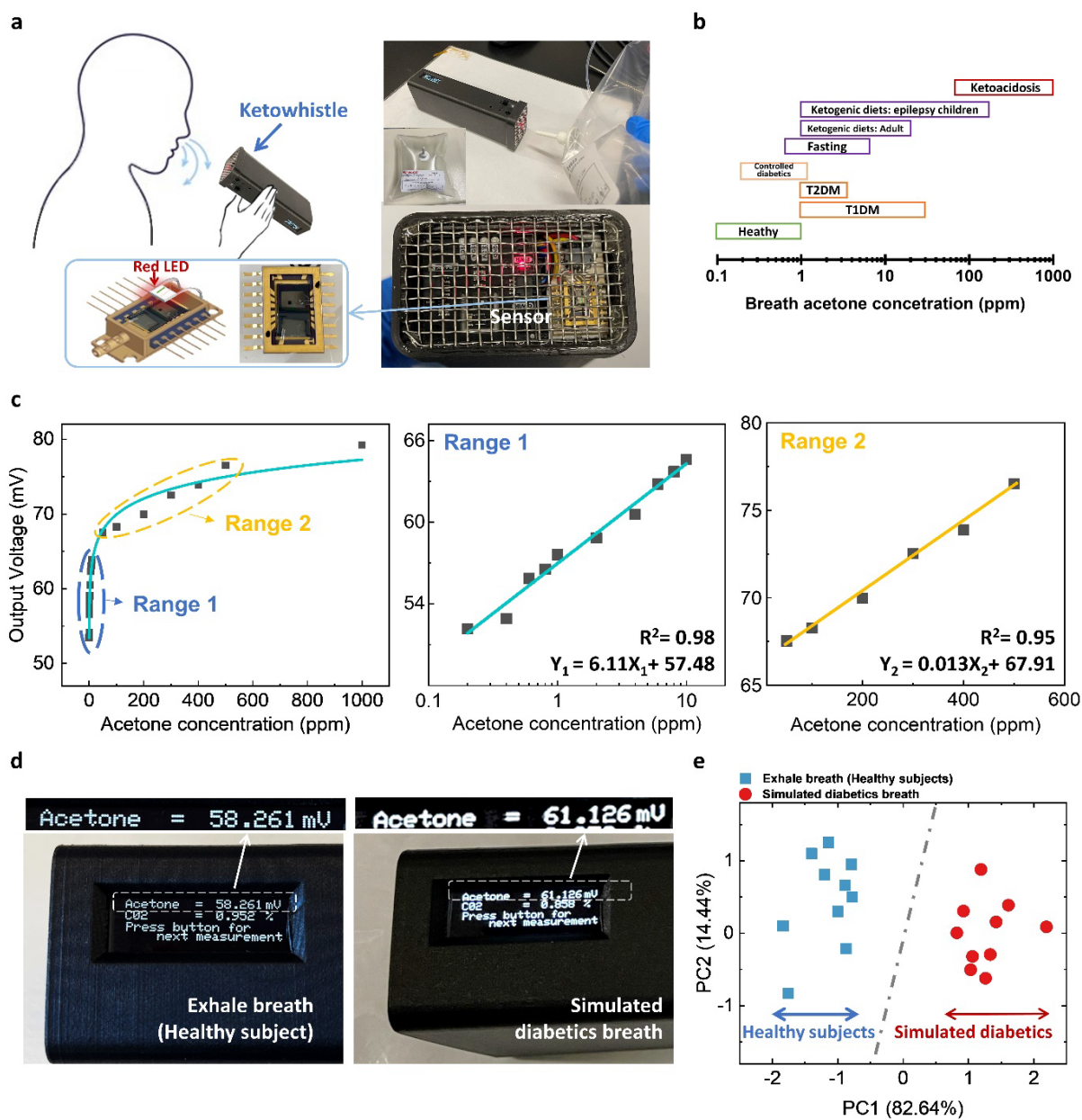

**Figure 5. The Ketowhistle for simulated breath testing. a**, Pictures of the portable Ketowhistle breath testing prototype using Tedlar bags for collection of simulated breath samples. **b**, Breath acetone spectrum showing the ranges of breath acetone concentration corresponding to various physiological states and ketosis ranges.[31c] **c**, Ketowhistle calibration for acetone concentration ranging from 0.1-1000 ppm to cover the entire breath acetone spectrum **d**, OLED screen displaying the response of an exhaled breath sample of a healthy subject (left) and a simulated diabetic breath sample (right). **e**, PCA for healthy exhaled breath samples and simulated diabetic breath samples.

For the exhaled breath test, Tedlar bags (PVDF 0.6 L, Sigma-Aldrich) suitable for VOC sampling were used to collect the breath samples from healthy subjects and simulated diabetic breath,[33] as shown by the inset in Figure 5a. According to Figure 5b, exhaled breath in simulated diabetic patients contains acetone gas with the concentration of 1.5 - 20 ppm. For the exhaled breath test, a concentration of simulated breath of diabetic patients was adjusted to ~ 5 ppm by filling the Tedlar bags ~50 % (v) of healthy exhaled breath and ~50 % of 10 ppm acetone in simulated air. The measurement result captured by the Ketowhistle is displayed on an OLED screen, as shown in Figure 5d. Principal component analysis (PCA) was applied for analysing the output voltage-dependent breath test results in Figure 5e,[34] where the distinction between the 10 simulated diabetic breath samples and 10 healthy subjects was clearly categorized into two distinguishable clusters without any overlap. The stability and reproducibility of the Ketowhistle prototype have been further demonstrated by repeated breath tests from the same person for an extended period (SI, Figure S11), indicating excellent consistency essential for clinical applications.

For the future work, the next step is for the chitosan/Pt/InP NW sensor based Ketowhistle prototype to be assessed in pilot clinical studies, including persons living with type 1 (well and during episodes of DKA), type 2 diabetes (including those using SGLT2 inhibitor medications), and overweight/obese persons using ketogenic diets in weight loss programs. To enhance device performance, the chitosan functionalization layer can be further modified to improve NW coverage by changing its concentration (by dilution) and/or composition.[17, 27] Guided by this study, various functionalization materials with functional groups specific to other breath VOCs are also possible to realize a variety of NW array-based multiplex breath sensor devices.

4. Conclusions

In this work, a novel, high-performance, non-invasive, self-powered, chitosan/Pt/InP nanowire acetone sensor has been successfully demonstrated for the first time. The sensor device exhibits superior sensitivity with an ultra-wide acetone sensing dynamic range, from the concentration of sub-ppb level up to >110,000 ppm, as well as high selectivity against other VOCs especially gaseous ketones. The sensing mechanism of this device is thoroughly investigated by controlled experiments and numerical simulations, revealing an oxygen-facilitated two-step charge transfer process for acetone reduction, enabled by chitosan functionalization and Pt/InP Schottky contact. Finally, by incorporating the nanowire sensor into a portable breath testing prototype, the Ketowhistle, we showed their immediate potential for non-invasive ketone testing and monitoring for persons living with diabetes, in particular for DKA prevention. This study also provides an exciting pathway towards future reliable breath analysis for diabetes diagnosis, management and beyond.

## 5. Methods

*Bottom-up growth of InP NW arrays by SA-MOCVD*

1) Substrate processing

Firstly, a 30 nm-thick $SiO_2$ layer was deposited on an n-doped (Si) (111)A InP substrate (1-10×10$^{18}$ cm$^{-3}$) by plasma enhanced chemical vapor deposition (PECVD) at 300 °C. A negative e-beam resist, AR6200.09 was spin-coated (step 1: 500 rpm for 5 s; step 2: 2000 rpm for 60 s) on the $SiO_2$ layer and baked at 150 °C for 1 min on the hotplate. Then, the resist was exposed to form a 400 × 400 μm hexagonal dot array pattern by a Raith 150 electron beam lithography system. After the development of resist, oxygen plasma (300 W, 2 min, 300 sccm $O_2$ flow) was used to remove the resist residues in the patterned area. The exposed pattern was etched by reactive ion etching to remove the $SiO_2$ layer on the InP substrate (20 sccm $CHF_3$, RF power: 20 W, 4.5 min). The exposed InP surface was trim etched using 10 % $H_2O_2$ (2 min) and 10 % $H_3PO_4$ (2 min) repeated sequentially for 5 times. After trim etching, the sample was immediately transferred into the MOCVD reactor for NW growth.

2) InP NW growth

The InP NW arrays were grown with an AIXTRON 200/4 MOCVD reactor, operating at a base pressure of 100 mbar, using $H_2$ as a carrier gas with a total flow of 14.5 L/min. Trimethylindium (TMIn) and phosphine ($PH_3$) were used as group III and group V precursors, respectively. Molar fractions of TMIn and $PH_3$ were set at 9.38×10$^{-6}$ mol/min and 7.59×10$^{-4}$ mol/min, respectively, corresponding to a V/III ratio of 80. All samples were hot baked at 750 °C for 10 min under a $PH_3$ protective flow and undoped InP NW (i-InP) arrays were grown for 4 min at 730 °C. For the growth of n-doped NW, silane ($SiH_4$) was introduced with all the other parameters kept the same as those used for the undoped InP NWs.

*Top-down NW array fabrication*

The top-down etching method for InP nanowire array fabrication (Extended Data Fig. 1a) has been reported in our previous study,[25] which includes the following steps: a) PECVD deposition of ~200 nm $SiO_2$ onto an undoped (n-type) InP wafer ((1-10) × 10$^{15}$ cm$^{-3}$); b) EBL patterning; c) Deposition of 70 nm Ni on the top of the EBL patterns by e-beam evaporation followed by lift-off in ZEP remover (Kirsten Hackenbroich) for 3 hours; (d) Inductively Coupled Plasma - Reactive Ion Etching (ICP-RIE) with fluorine (ICP-F) gas to etch the exposed $SiO_2$ layer to form a Ni/$SiO_2$ bilayer as the mask for subsequent InP etching; and (e) InP etching with Ni/$SiO_2$ mask by ICP-RIE with chlorine gas (ICP-CL).

The NW diameter was determined by the mask and the ICP etching process. The smallest mask size that can be achieved is ~ 40-50 nm through EBL and Ni deposition. The SEM images in Extended Data Fig. 1d and Fig S4 (SI) show that the fabricated NWs exhibit large tapering due to the etching of the exposed NW sidewalls during the deep etching of the substrate, leading to a gradual change in the NW diameter from top (~30 nm) to bottom (~130 nm). Another limitation of the top-down approach is the shorter NW length due to the etching selectivity of the mask and InP substrate, as the Ni/$SiO_2$ mask for ICP-RIE etching will be slowly removed during the etching process. The average length of top-down etched NW is ~2 μm, considerably shorter than that of the bottom-up grown NWs (> 4 μm).

*Acetone sensor fabrication*

To fabricate a chemiresistive sensor based on the InP NW array, SU8-5 photoresist (Kirsten Hackenbroich) was spin-coated to cover the entire NW array. To prevent the NW from breaking up during the fabrication process, a low spin speed of 1000 rpm was applied, followed by a two-step baking process at 65 °C and 95 °C for two minutes each. Then, the SU8-5 film was etched by a barrel-etcher (PVA Tepla Gigabatch 310 M) with an $O_2$ flow rate of 300 sccm and a power of 500 W to expose the top ~500 nm of the NWs for subsequent electrical contact. Then, the sample was flood exposed under UV illumination and baked at 150 °C to solidify the SU8-5 film. This SU8-5 layer is used to electrically isolate the top of the NW from the InP substrate, as well as to provide mechanical support to the NWs. E-beam evaporator was then used to deposit 60 nm Pt layer on the nanowires for top contact, and the samples were mounted on a special holder to enable the tilted-angle deposition. This tilted-angle deposition method was applied to interconnect all NWs for electrical signal extraction and partially expose the NW surface for gas sensing measurement. Finally, the chitosan acetic acid aqueous solution with a concentration of 2.5 % (diluted at a 1:10 ratio from the saturated solution with a concentration of 28 %) was drop-casted on the Pt-deposited NW followed by baking on a hotplate at 50 °C to remove the solvent.

*Dark/light current-voltage characterization*

The dark/light current-voltage (I-V) characteristics of the chitosan/Pt/InP NW array device were characterized by a Keysight 2900 source/measure unit under 1 Sun @AM1.5 condition (solar simulator, 100 mW/cm$^2$). The current-voltage (I-V) curve of the device shows a rectification effect (Fig. 2a, b), indicating a typical Schottky contact formed between InP NW and Pt.[35] A photovoltaic effect is observed under the light illumination from solar simulator with a short circuit current ($I_{SC}$) of 136 nA and an open circuit voltage ($V_{OC}$) of 80 mV, which is the foundation of further self-powered sensing operation[36] with the $I_{SC}$ acting as the self-powered sensing signal.

*Acetone sensing measurements*

The gas sensing performance was measured by a custom-made sensing setup consisting of a Linkam chamber with a sample stage and Au probes, mass flow controllers (MFCs Bronkhorst), a solar simulator, and gas cylinders as shown in Fig. 2a. For gas sensing measurement, the carrier gas is the simulated air with a volume ratio of N$_2$ to O$_2$ at 4 ($V_{N2}/V_{O2}$ = ~ 4, N$_2$ and O$_2$, BOC gas). The gas flow rate was controlled by MFCs, while the total gas flow rate was kept at 1 L/min for ppm level concentrations and 0.5 L/min for sub-ppm level concentration measurements. For the measurement of analyte gas, the target VOCs (ethanol, 9.91 ppm in N$_2$, Coregas; NO$_2$, 10.1 ppm in N$_2$, Coregas; methanol, 10 ppm in N$_2$, BOC gas; acetone, 10 ppm in N$_2$, BOC gas) was diluted with the simulated air to the desired concentration before being passing through the chamber.

Sensing response (R) quantifies the ratio of sensing signal change before and after the exposure to an analyte, which can be calculated by the following equation:

$$R = \frac{(I_{gas} - I_0)}{I_0} \times 100\%, \tag{1}$$

where $I_0$ and $I_{gas}$ denote the $I_{SC}$ measured in simulated air and upon acetone gas exposure, respectively.

Sensitivity (S): generally, S is defined as the minimum perturbation of physical parameters that will create a detectable output change. S can be obtained from the slope of the response vs. concentration curve S = R/C (%/ppb), where C donates the concentration.

Limitation of detection (LOD): LOD is defined as the lowest concentration or amount of a substance that can be reliably detected and distinguished from the background at three times the standard deviation of sensing noise ($SD^2_{noise}$), i.e., LOD = 3×($SD^2_{noise}/L$), where $L$ is the slope of the calibration result and also the sensitivity as mentioned in the main text.[28a] The $SD^2_{noise}$ of both devices were obtained from the standard deviation of 100 consecutive data points of the baseline. For the bottom-up NW sensor (b-chitosan/InP/NW sensor), the LOD is 0.18 ppb, with calibration Y= 0.48 (%/ppb) X+0.71 (Fig. 2e) and the $SD^2_{noise,b}$ = 0.27; the LOD of the top-down NW sensor (t-chitosan/InP/NW sensor) is 82 ppb with $SD^2_{noise,t}$ = 0.31, calibration Y= 6.61 (%/ppm) X+0.39 (Extended Data Fig. 2e).

Humidity test: Water vapor is produced by the same bubbler setup as above, where the organic solvent is replaced by DI water in the bubbler. The water vapor is injected into the gas chamber with a constant $N_2$ flow rate during the whole sensing process, and the relative humidity (RH) is controlled by the $N_2$ flow rate, as shown in Table S1.

High-concentration vapor is produced by a bubbler setup with a tube inlet for $N_2$ to flow into the relevant solvent, evaporating saturated vapor to the gas sensing chamber (Fig. 2a). To stabilize the temperature, the bubbler is placed in a water bath maintained at 25 °C during the whole sensing measurement process. The vapor concentration ($C_{con}$) was calculated by following equations:[37]

$$F_{output} = \left(\frac{P}{P_o - P}\right) F_{carrier} = \alpha F_{carrier}, \quad (2)$$

$$C_{con} = \frac{10^6 F_{output}}{F_{dilute} + F_{carrier} + F_{output}} = \frac{10^6 \alpha}{(F_{dilute}/F_{carrier}) + 1 + \alpha}, \quad (3)$$

where $F_{output}$, $F_{carrier}$, and $F_{dilute}$ are the output, carrier, and dilute flow rate of acetone vapor from the bubbler, respectively, in standard cubic centimetre per minute (sccm), and α is the vapor pickup efficiency defined as the relative ratio of the output sample flow rate to the carrier flow. $P_o$ is the outlet pressure in the bubbler, defined by $P_o = P_i + P_{th}$, where $P_i$ is the inlet pressure to the bubbler (1.6 bar), and $P_{th}$ is the thermodynamic vapor pressure of the analyte sample. When the carrier gas is completely saturated with the analyte vapor, $P_{th}$ becomes the saturated vapor pressure, $P_s$, which can be calculated from the empirical Antoine equation[38]:

$$log_{10} P_s = A - \frac{B}{C+T}, \quad (4)$$

where $T$ is the temperature (in °C or K according to the value of C), and A, B, and C are component-specific constants of 4.42, 1312.25, and -32.45, respectively, for acetone at 25 °C. The calculated $P_s$ of acetone is 30.6 kPa (298 K) (https://webbook.nist.gov/cgi/cbook.cgi?ID=C67641). With different dilution ratios of the simulated air, a series of high acetone concentrations from 0.57 % (5700 ppm) to 16.05% (160,500 ppm) can be obtained through this setup (Table 1).

Table 1. Parameters for calculating the acetone concentration produced from the bubbler.

| $F_{carrier}$ (sccm) | $F_{vapor}$ (sccm) | $F_{dilution}$ (sccm) | α | Vapour Conc (ppm) | Vapour Conc (%) |
|---|---|---|---|---|---|
| 1000 | 191.21 | 0 | 0.15 | 160516.63 | 16.05 |
| 830 | 158.70 | 170 | 0.15 | 136966.31 | 13.70 |
| 660 | 126.20 | 340 | 0.15 | 112056.53 | 11.21 |
| 500 | 95.60 | 500 | 0.15 | 87261.80 | 8.73 |
| 330 | 63.10 | 670 | 0.15 | 59353.75 | 5.94 |
| 170 | 32.51 | 830 | 0.15 | 31482.16 | 3.15 |
| 130 | 24.86 | 870 | 0.15 | 24254.26 | 2.43 |
| 100 | 19.12 | 900 | 0.15 | 18762.13 | 1.88 |
| 66 | 12.62 | 973 | 0.15 | 12462.51 | 1.25 |
| 30 | 5.74 | 970 | 0.15 | 5703.55 | 0.57 |

Gas sensing measurement under $O_2$ eliminated condition: This condition was realized by using pure $N_2$ flow flashing the gas sensing setup for more than one hour to minimize the $O_2$ residue in gas sensing setup and set pure $N_2$ as the carrier gas during the sensing test.

*Breath test with Ketowhistle*

Exhaled breath testing by the Ketowhistle was performed using Tedlar bags (PVDF 0.6 L, Sigma-Aldrich) with the Push/Pull Lock Valve (Fig. 5a). To simulate diabetic breath, the Tedlar bags were filled with ~50 % (v) of healthy exhaled breath mixed with ~50 % of 10 ppm acetone. This volume composition was achieved by pumping in acetone gas with MFCs at 1 L/min flow rate for 18s to fill the Tedlar bag with 0.3 L acetone gas, and equal volume of exhale breath.

Acknowledgements

The authors would like to acknowledge the Australian Research Council for financial support and the Australian National Fabrication Facility (ACT node) for facility support. This research was also funded by and has been delivered in partnership with Our Health in Our Hands (OHIOH), a strategic initiative of the Australian National University, which aims to transform healthcare by developing new personalized health technologies and solutions in collaboration with patients, clinicians, and health care providers. A.T. gratefully acknowledges the support of the Australian Research Council for a Future Fellowship (FT200100939) and Discovery grant DP190101864. A.T. also acknowledges financial support from National Intelligence and Security Discovery Research Grants NS210100083. S.W. thanks the China Scholarship Council and the Australian National University for scholarship support. The authors also thank Dennis Gibson, Michael Blacksell, Steve Marshall from the Electronics Unit of the Research School of Physics, the Australian National University for the Ketowhistle development and production.

Disclaimers

The authors have filed a provisional patent for this work.

# Supporting Information

Nanowire Array Breath Acetone Sensor for Diabetes Monitoring

*Shiyu Wei, Zhe Li, Krishnan Murugappan, Ziyuan Li, Fanlu Zhang, Mykhaylo Lysevych, Hark Hoe Tan, Chennupati Jagadish, Buddini. I Karawdeniya[*], Christopher Nolan, Antonio Tricoli[*], and Lan Fu[*]*

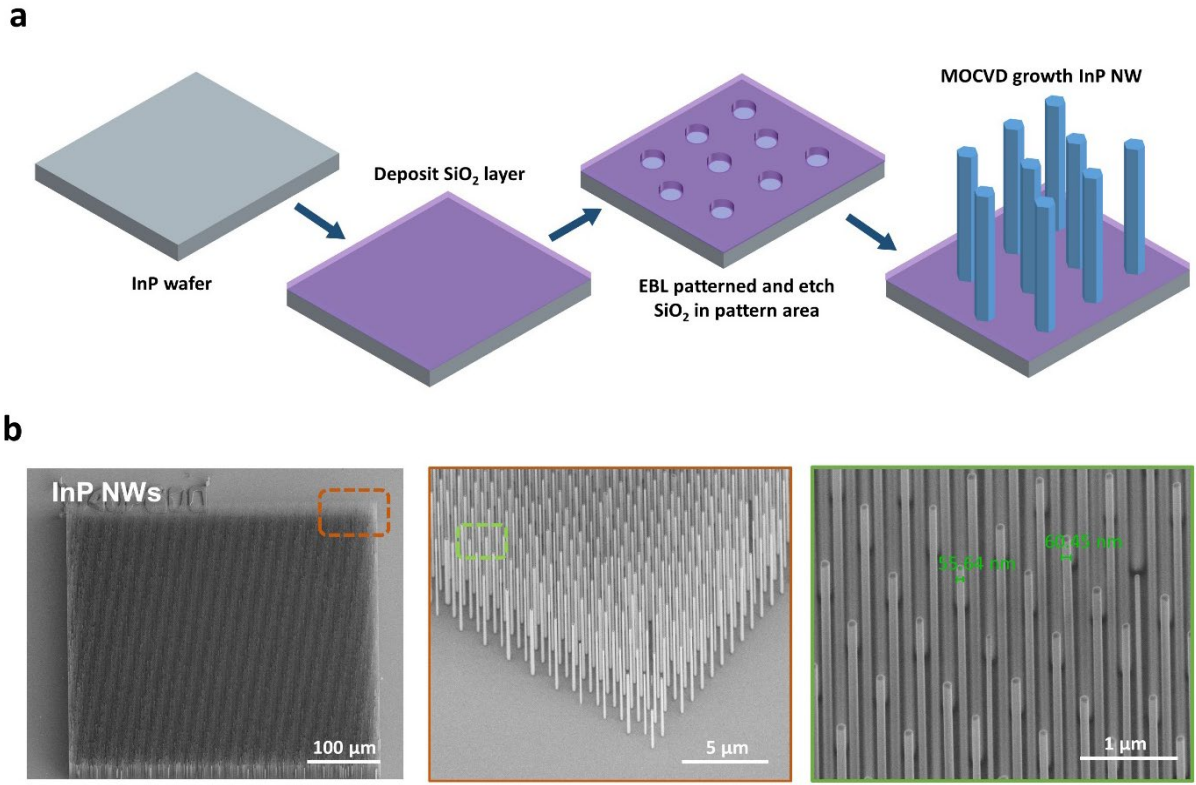

Figure S1. **a**, Schematics of selective area metal-organic chemical vapor deposition (SA-MOCVD) nanowire (NW) array growth. This method starts with substrate preparation process including the deposition of a 30 nm SiO₂ layer on InP substrate followed by electron beam lithography (EBL) patterning and reactive ion etching to form the hexagonal dot array pattern for NW array growth. **b**, The scanning electron microscope (SEM) images of the SA-MOCVD grown InP NW arrays under different magnification to characterize the overall NWs array morphology, detailed NW structure imaging (orange box) and diameter measurement (green box).

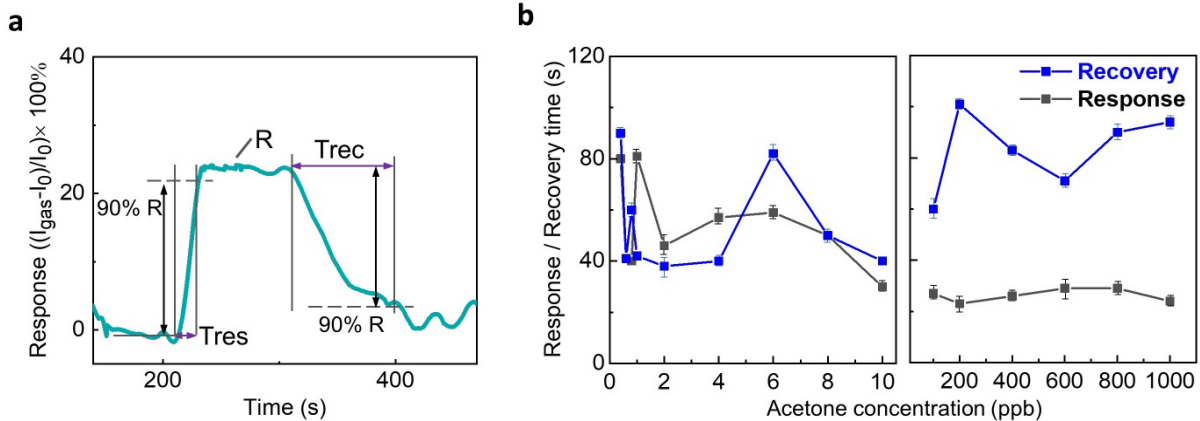

Figure S2. **a**, Calculation of the response and recovery time from the time dependent acetone sensing measurement at 200 ppb concentration. **b**, The calculated acetone sensing response and recovery time corresponding to the acetone concentration range of 0.4-10 ppb and 100-1000 ppb in Fig 2d, e, with the standard deviation as the error bars obtained by 10 cycles of sensing measurements.

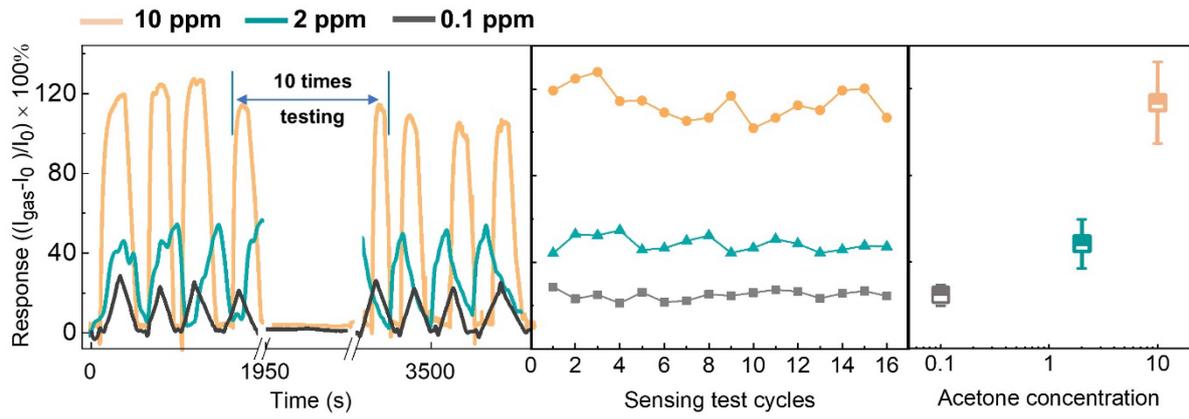

Figure S3. The reproducibility measurement of b-chitosan/Pt/InP NW sensor by repeating 16 cycles of the acetone sensing measurement.

Table S1. The humidity test at different RH controlled by flow rates of the mass flow controller

| RH (%) | N$_2$ to the bubbler (L·min$^{-1}$) | Simulated air (L·min$^{-1}$) |
|---|---|---|
| 20 | 0.2 | 0.8 |
| 50 | 0.5 | 0.5 |
| 65 | 0.65 | 0.35 |

Table S2. Comparison of sensing performance of different chemiresistive acetone sensors reported in recent literature.

| Materials | Working temperature (°C) | Response (concentration) | Limit of detection | Response/ recover time (s) | Ref |
|---|---|---|---|---|---|
| Chitosan/Pt/InP NW | 25 | 48% (1 ppm) | 0.4 ppb | 25 / 39 | This work |
| t-chitosan/InP NW | 25 | 6.5% (1 ppm) | 82 ppb | 5 / 6 | This work |
| NiO/NiWO$_4$/WO$_3$ (p–p–n) nanowire | 400 | 29 (30 ppm) | 0.7 ppm | - | [1] |
| Pt/WO$_3$ hemitubes | 350 | 4.11 (2 ppm) | 120 ppb | - | [2] |
| Si/WO$_3$ | 135 | 5.4 (1 ppm) | 50 ppb | - | [3] |
| Rh/SnO$_2$ nanofibers | 200 | 2.7 (1 ppm) | - | 2 / 64 | [4] |
| chitosan–Pt SnO$_2$ nanofibers | 350 | 2.9 (5 ppb) | 5 ppb | 12 / 84 | [5] |
| Au- ZnO/Ag core-shell films | 150 | 43% (0.5 ppm) | 0.5 ppm | 45 / 160 | [6] |
| Na/ZnO Nanoflowers | 25, (UV light, 5 mW·cm$^{-2}$) | 1.51 (1 ppm) | 0.2 ppm | 18 / 63 | [7] |
| Zn$_3$N$_2$ / ZnO | 200 | 5.1 (1 ppm) | 0.07 ppm | 15 / 27 | [8] |
| In$_2$O$_3$ fibers | 275 | 24 (1ppm) | 0.3 ppm | - | [9] |
| poly[(9,9-dioctylfluorenyl-2,7-diyl)-co-(4,4-(N-(4-sec-butylphenyl) diphenylamine)] | 25 | 13.3% (1 ppm) | 0.3 ppm | - | [10] |
| Chitosan thin film | 25 | 12.3% (1 ppm) | 0.1 ppm | - | [11] |

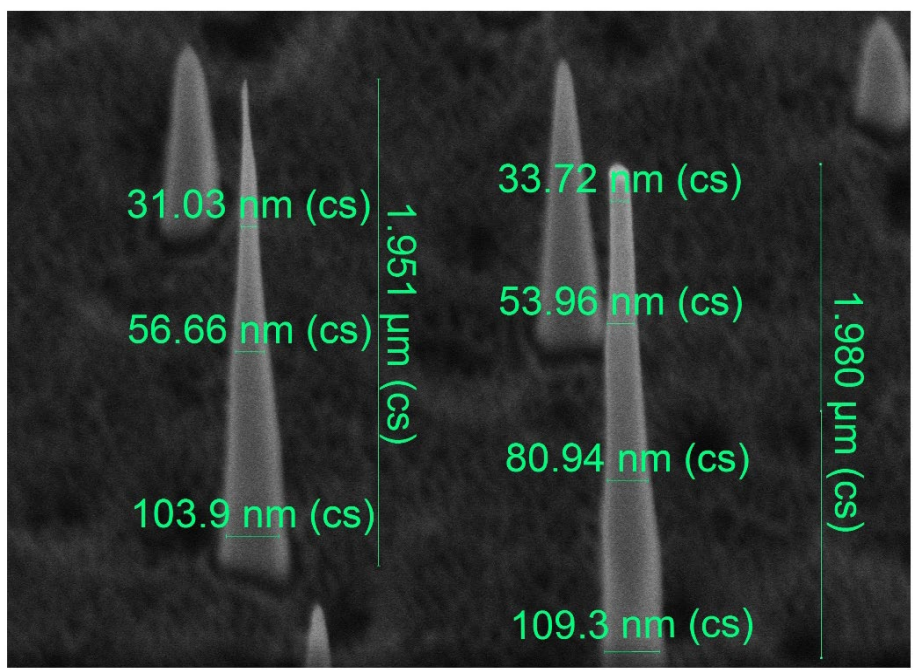

Figure S4. The SEM image of the top-down etched InP NWs with the diameter and length measurements. The NWs exhibit slightly different morphology, i.e., size, shape and surface roughness from the bottom-up grown NWs shown in Fig. S1.

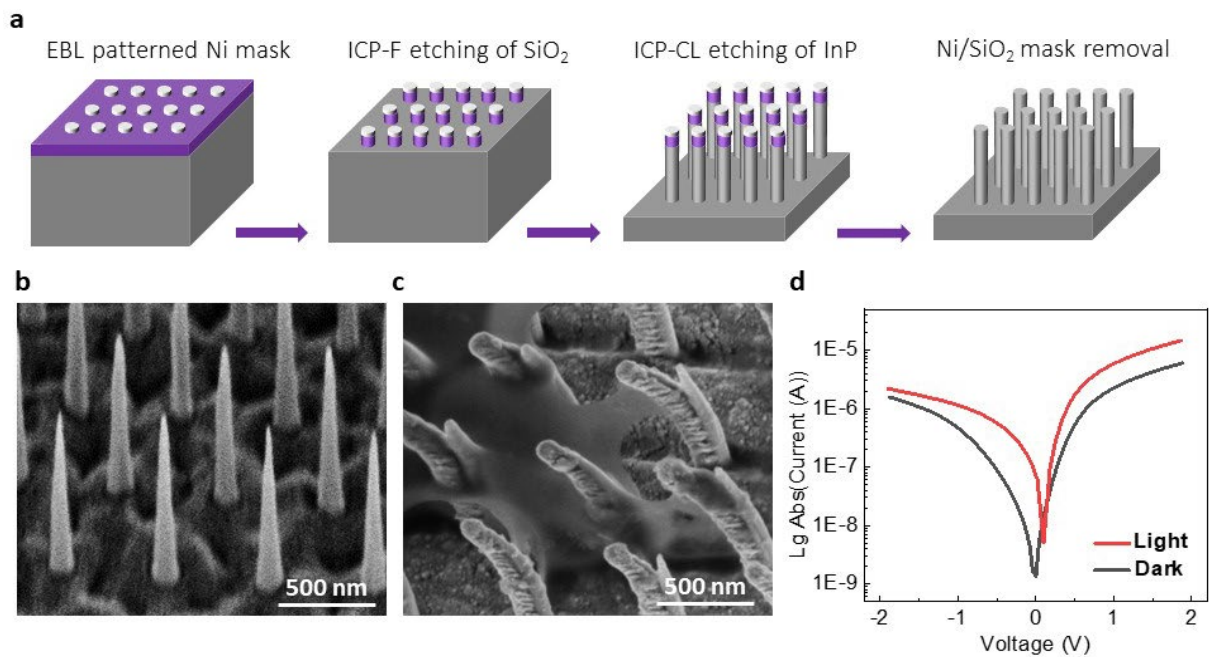

Figure S5. The top-down etched NW array and device (t-chitosan/Pt/InP) fabrication and electric measurement. **a,** Schematics of top-down etching approach for InP NW array fabrication: EBL patterned Ni (thickness 70 nm, diameter 40 nm) mask on PECVD deposited $SiO_2$ (thickness 600 nm), which was pre-deposited on undoped InP wafer ($(1-10) \times 10^{15}$ cm$^{-3}$); ICP-F etching of $SiO_2$ layer to form the $SiO_2$/Ni mask; ICP-CL etching of the InP wafer with the $SiO_2$/Ni mask; $SiO_2$/Ni mask removal by 10% HF solution. SEM images of **b**, InP NWs after ICP-CL etching, and **c**, after Pt and chitosan deposition. **d**, I-V measurement of the t-chitosan/Pt/InP NW sensor under the dark/light condition with solar simulator @AM 1.5, 100 mW/cm$^2$.

Due to the small and non-uniform NW diameter and possible top-down etching-induced surface damage, the top-down NW device exhibits a weaker photovoltaic effect with an $I_{SC}$ of 61 nA and $V_{OC}$ of 99 mV. Nevertheless, it is sufficient for self-powered operation, producing a selective, stable, and fast response highly desirable for breath testing.

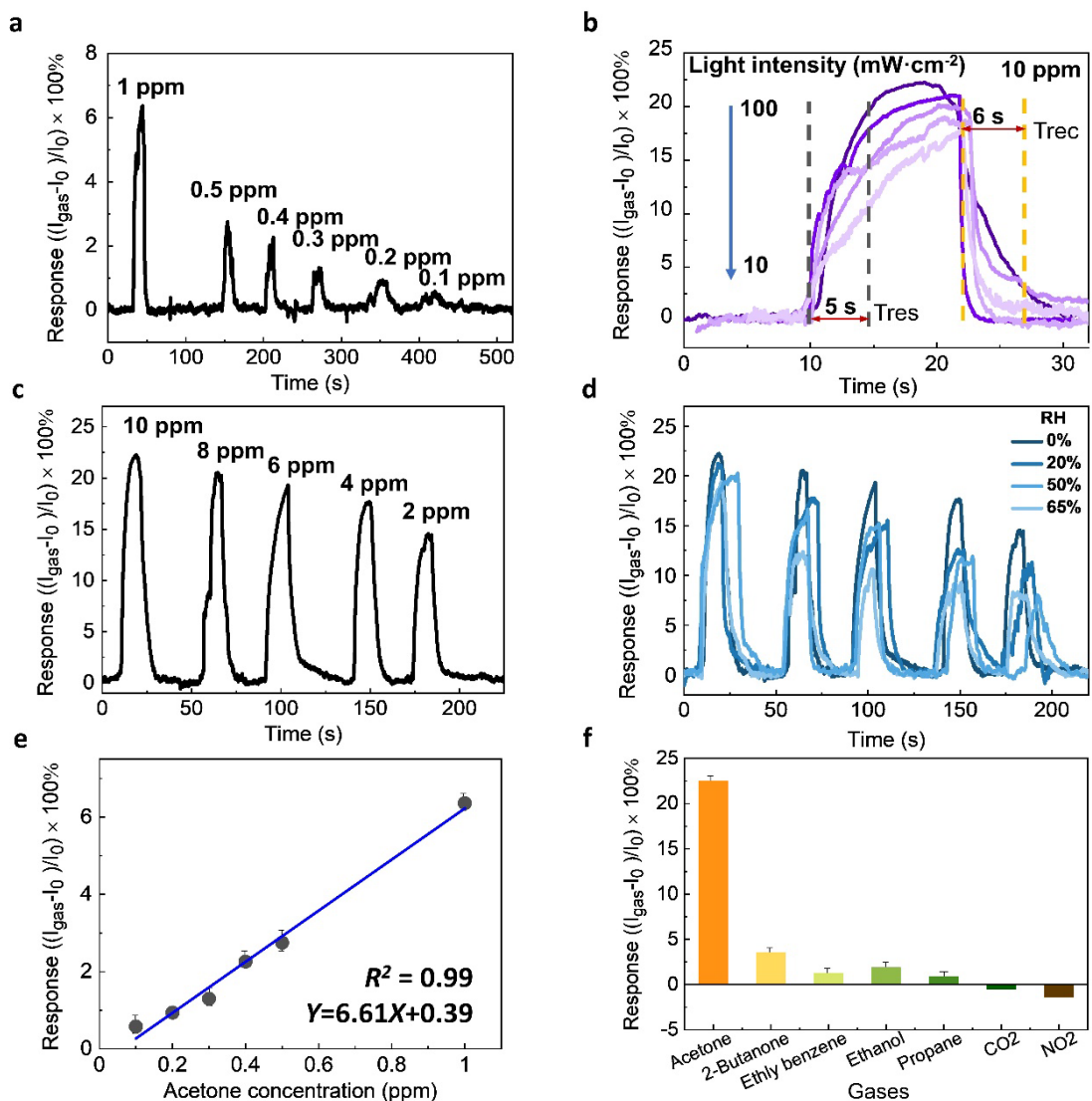

Figure S6. Acetone sensing performance of the top-down method fabricated NW sensor (t-chitosan/Pt/InP). **a**, Time-dependent sensing response measured for an acetone concentration of 0.1-1 ppm. **b**, Acetone sensing response under different light illumination intensities from 100 to 10 mW·cm$^{-2}$, with the response and recovery times of the response curve under 100 mW·cm$^{-2}$ light illumination being indicated. **c**, Time-dependent sensing response curve with an acetone concentration range of 2-10 ppm. **d**, Time-dependent sensing response to the acetone concentration of 2-10 ppm under the relative humidity (RH) levels of 0%, 20%, 50%, 65%. **e**, Sensor response vs concentration curve with linear fitting. **f**, Gas sensing selectivity measurement to compare the response to 10 ppm acetone, 2-butanone, ethyl benzene, ethanol, propane, NO$_2$, and 10% CO$_2$. The error bars in **e**, **f** indicate the standard deviation obtained from 10 cycles of sensing measurements.

The performance comparison between b-chitosan/Pt/InP NW and t-chitosan/Pt/InP NW acetone sensor is summarized in Table S2, indicating that the bottom-up NW sensor is more sensitive than the top-down sensor, whereas the latter has a faster response speed. This may be ascribed to the less ideal NW morphology due to top-down etching, i.e., the tapered and bent NWs with a rough surface. Such morphology causes less chitosan coverage, and slightly degraded electrical property. Nevertheless, the performance of the top-down NW sensor is adequate for breath acetone sensing, as confirmed by the successful demonstration of Ketowhistle prototype; in particular, the fast response speed is highly desirable for breath-sensing applications.

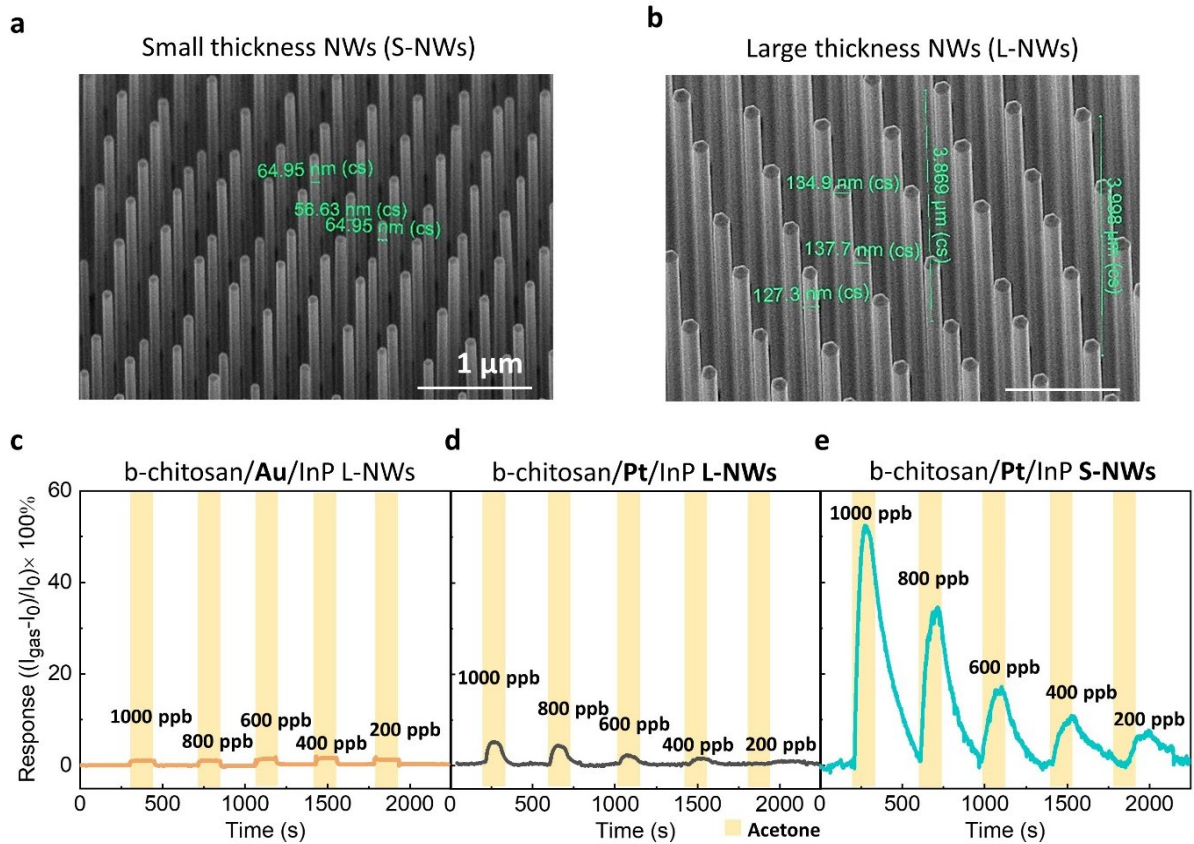

Figure S7. **a**, **b**, SEM image of small (50-60 nm) and large (120-140 nm) diameter NW array by bottom-up SA-MOVPE techniques, respectively. The time-dependent sensing response measured from: **c**, large diameter NW array sensor with Au electrode, i.e., b-chitosan/Au/InP L-NWs; **d** large diameter NW array sensor with Pt electrode, i.e., b-chitosan/Pt/InP L-NWs; **e**, small diameter NW array sensor with Pt electrode, i.e., b-chitosan/Pt/InP S-NWs.

To investigate the effect of NW diameter on acetone sensitivity, larger diameter NW array was grown (Fig. S5b).[26] As shown in Fig. S5c and d, the devices only produced a small response (< 10 %) to acetone with the concentration up to 1000 ppb. By reducing the NW diameter to ~50-60 nm, a significant sensitivity enhancement was observed (Fig. S5e). As Pt was replaced by Au contact which has a smaller work function, the effectiveness of the Schottky contact was much reduced, causing a much-increased base-line current and a non-specific response to acetone (Fig. S5c). This can be explained by the strong influence of Pt Schottky contact on thin NWs, which generates a strong built-in electric field that effectively modulate the electron concentration at the NW surface facilitating the $O_2$ ionization and acetone reduction process.

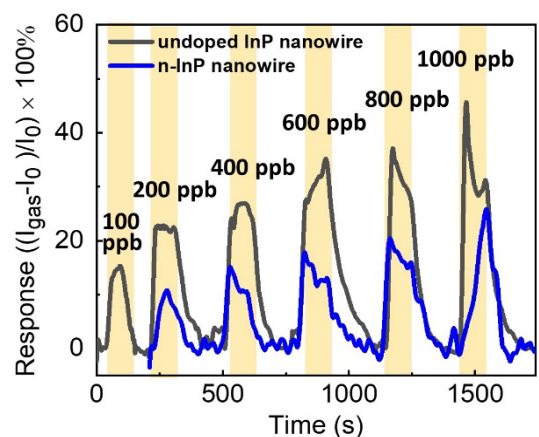

Figure S8. The time-dependent acetone sensing measurement with the chitosan modified InP NW sensors which are undoped and n-doped sensor at zero bias, respectively. The device fabricated from n-doped InP NW (doping concentration ~$3\times10^{18}$ cm$^{-3}$) showed a decreased response compared to the undoped InP NW sample due to the larger baseline current.

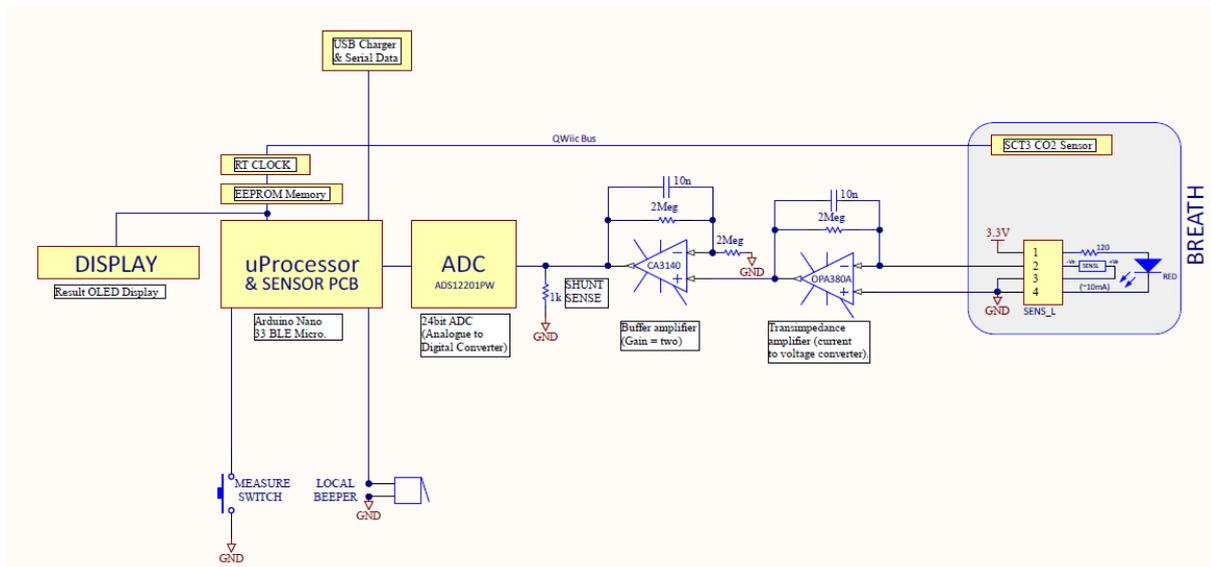

Figure S9. Block diagram of the electrical circuit in Ketowhistle. The acetone sensor is located at the "Breath" frame and powered by a low power red LED. The output signal is captured by measuring the voltage of a 2M load resistor in series with the amplified sensor current signal. The actual current from the acetone sensor can be calculated as: current = output voltage / (4 × $10^6$ $\Omega$).

Table S3. The parameters of the electrical elements in the block diagram shown in Fig S7.

| Comment | Description |
| --- | --- |
| Op Amp (CA3140) | FET Operational Amplifier (Gain = two) |
| Op Amp (OPA380A) | Transimpedance amplifier (current to voltage converter) |
| Cap | Capacitor |
| Red | Typical Infrared GaAs LED (Digikey, LS R976-NR-1), power = 3.2 mW, peak wavelength = 645 nm |
| Beeper | Magnetic Transducer Buzzer |
| SWITCH | Switch |
| SENS_L | Header, 4-Pin |
| Res1 | Resistor 2M ohm |
| SENSL | Resistor |

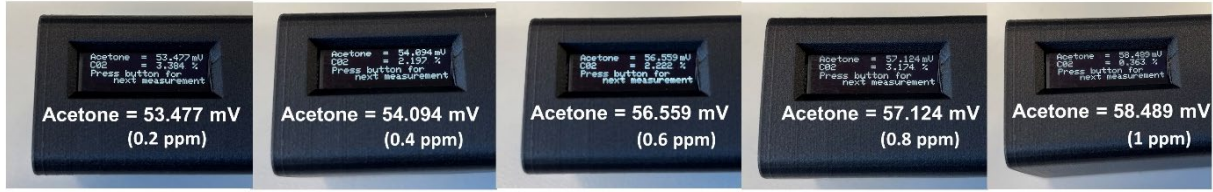
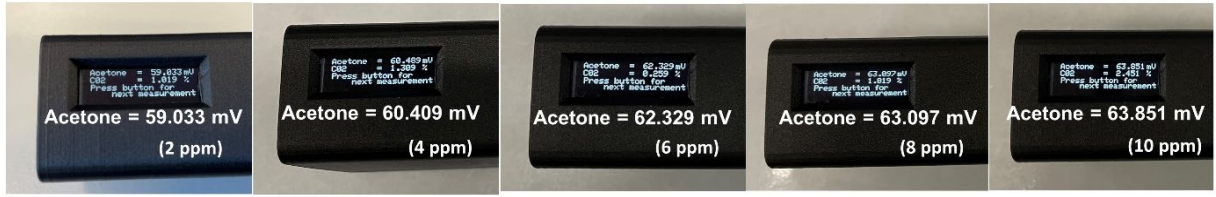
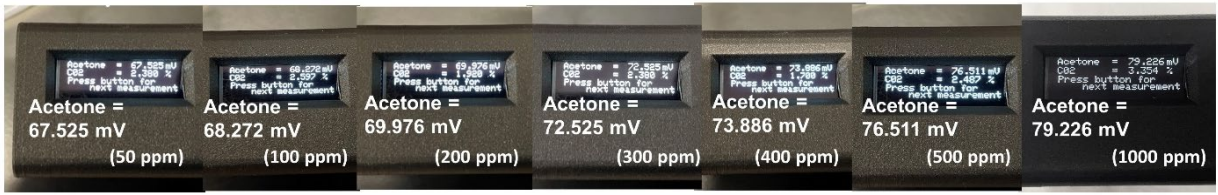

Figure S10. Recorded pictures for ketone whistle calibration for low (0.1 – 1 ppm), medium (1 – 10 ppm), and high (50 - 1000 ppm) acetone concentration range, respectively.

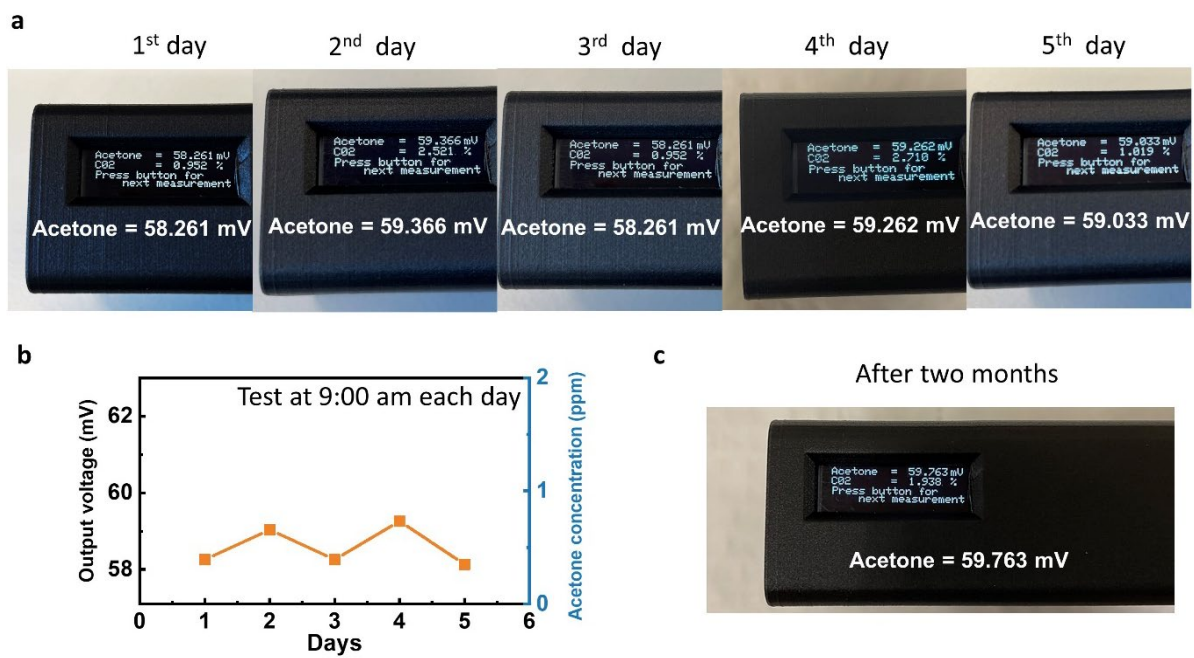

Figure S11. **a**, The Ketowhistle breath testing result recording from the same person at 9:00 am for five consecutive days and **b**, the corresponding acetone concentration based on the calibration data presented in Fig. 5c. **c**, The Ketowhistle breath testing result recorded from the same person after two months with the Ketowhistle stored in the ambient condition.